\documentclass[preprint,aps,a4paper,showpacs,showkeys,amsfonts,amsmath,amssymb]{revtex4}
\usepackage{graphicx}
\newcommand{\NN}{\mathcal{N}\/}
\begin{document}

\title{Mean Exit Time and Survival Probability within the CTRW Formalism}
\author{Miquel Montero}
\email[Corresponding author: ]{miquel.montero@ub.edu}
\author{Jaume Masoliver}
\affiliation{Departament de F\'{\i}sica Fonamental, Universitat de
Barcelona,  Diagonal 647, E-08028 Barcelona, Spain}

\begin{abstract}

An intense research on financial market microstructure is presently in progress. Continuous time random walks (CTRWs) are general models capable to capture the small-scale properties that high frequency data series show. The use of CTRW models in the analysis of financial problems is quite recent and their potentials have not been fully developed. Here we present two (closely related) applications of great interest in risk control. In the first place, we will review the problem of modelling the behaviour of the mean exit time (MET) of a process out of a given region of fixed size. The surveyed stochastic processes are the cumulative returns of asset prices. The link between the value of the MET and the timescale of the market fluctuations of a certain degree is crystal clear. In this sense, MET value may help, for instance, in deciding the optimal time horizon for the investment. The MET is, however, one among the statistics of a distribution of bigger interest: the survival probability (SP), the likelihood that after some lapse of time a process remains inside the given region without having crossed its boundaries. The final part of the article is devoted to the study of this quantity. Note that the use of SPs may outperform the standard ``Value at Risk" (VaR) method for two reasons: we can consider other market dynamics than the limited Wiener process and, even in this case, a risk level derived from the SP will ensure (within the desired quintile) that the quoted value of the portfolio will not leave the safety zone. We present some preliminary theoretical and applied results concerning this topic.

\end{abstract}
\date{\today}

\maketitle

\section{Introduction}

The continuous time random walk (CTRW) formalism was introduced four decades ago by Montroll and Weiss~\cite{montrollweiss}, as a natural extension of ordinary random walks (RWs). In
a (one dimensional) RW you can randomly move through a fixed grid either up or down, at regular time steps, whereas in a CTRW the size of the movements and specially the time lag between them are random. CTRWs have been successfully applied to a wide and diverse variety of physical phenomena over the years \cite{weissllibre}: transport in random media, random networks, self-organized criticality, earthquake modelling; and recently also to finance 
\cite{general1,general2,general4,general5,general6,ivanov,masoliver1,masoliver2,montero1,montero2}. In this latter context, the efforts have been mostly focused on the statistical properties of  the waiting time between successive transactions and the asset return at each transaction. Different studies in different markets are conceiving the idea that the empirical distributions of both random variables are compatible with an asymptotic fat tail behaviour \cite{general1,general2,general4,general5,general6,masoliver1,masoliver2,ivanov}. 

Within the CTRW formalism we have recently investigated the mean exit time (MET) of asset prices out of a given region for financial time series \cite{montero1, montero2}. In these articles we show that the MET follows a quadratic growth in terms of the interval width, both in small and large scales. We checked the persistence of this behaviour in time series from several markets, such the foreign exchange market, or the New York Stock Exchange (NYSE). The theoretical model used in these works was based on two-state chain Markovian processes. This model is able to both describe the quadratic scaling property observed for the MET and provide a mechanism that can incorporate asset peculiarities through return autocorrelations. 

One of the possible applications of the analysis of the MET in finance is in the field of risk control. There is a direct link between the value of the mean exit time out of a region, and the timescale of market fluctuations of a certain size. Therefore, its value may help, for instance, in deciding the minimal time horizon for an investment, the rotation rate of a portfolio, or even the value of stop-loss and stop-limit levels for a position. 

However, the mean exit time is only a statistic of a distribution with even bigger interest: the survival probability (SP), the probability that after some elapsed time a process remains inside the given region without having crossed its boundaries. This quantity may outperform the standard ``Value at Risk'' (VaR) method for two reasons: it could be based on market statistics different than the (unrealistic) Gaussian distribution, and it will ensure (within the desired quintile) that the market value of the portfolio will not leave the safe zone. 

The paper is organized as follows. In Sect.~\ref{Sect_Mean} we discuss the MET within the CTRW formalism,  under a meaningful set of simplifying assumptions. In Sect.~\ref{Sect_Markov} we relax some of the previous constrains in order to introduce some memory into the process. Section~\ref{Sect_SP} is devoted to the SP, its properties and its connections with the MET. In Sect.~\ref{Sect_Risk} we show in a practical situation how SP can be used in risk control. Conclusions are drawn in Sect.~\ref{Sect_Conclusions}.

\section{Extreme events within CTRW}
\label{Sect_Mean}

\begin{figure}[hbtp]
  \begin{center}
  \includegraphics[width=0.90\textwidth,keepaspectratio=true]{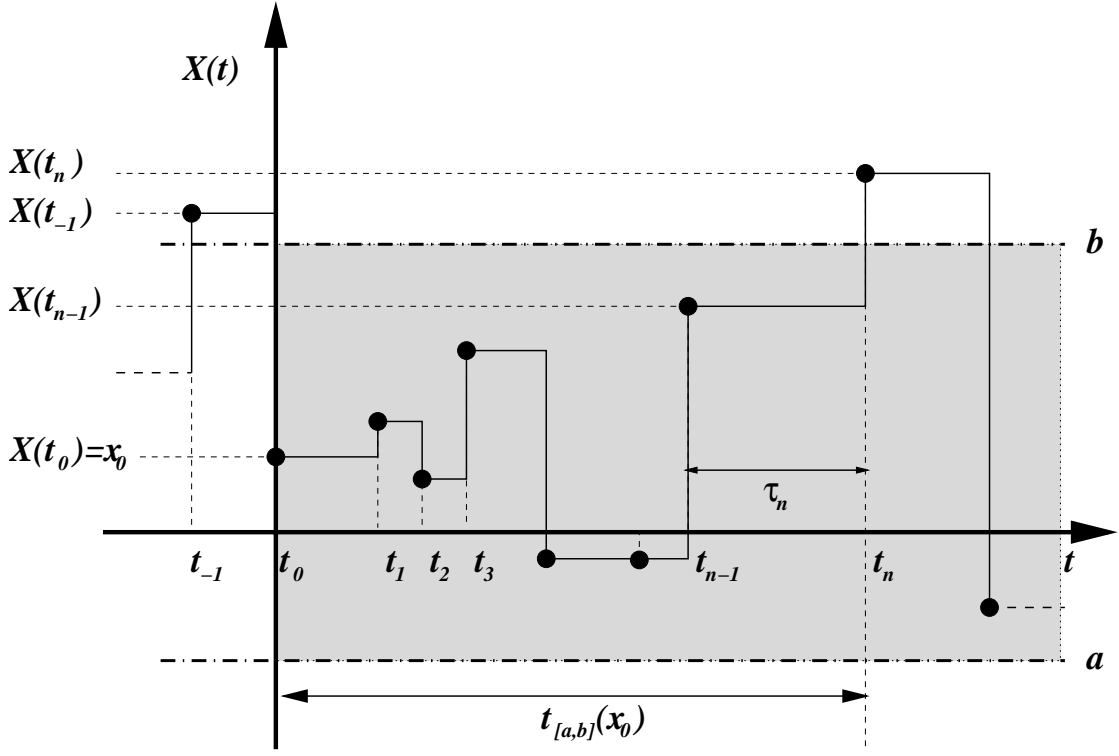} \caption{A sample trajectory of the $X(t)$ process along with the corresponding value of the random variable $t_{[a,b]}(x_0)$.} \label{model}
\end{center}
\end{figure}

In the most common version of the CTRW formalism a given random process $X(t)$ shows a series of random increments or jumps at random times $\cdots,t_{-1},t_0,t_1,t_2,\cdots,t_n,\cdots$ remaining  constant between these jumps. Therefore, after a given time interval $\tau_n=t_n-t_{n-1}$, the process experiences a random increment $\Delta X_n(\tau_n)=X(t_n)-X(t_{n-1})$ and the resulting trajectory consists of a series of steps as shown in Fig.~\ref{model}. Waiting times $\tau_n$ and random jumps $\Delta X_n(\tau_n)$ are described by their probability density functions (pdfs) which we will denote by $\psi(\tau)$ and $h(x)$ respectively. We refer the reader to Refs.~\cite{montrollweiss,weissllibre,general1,general2,general4,general5,general6,ivanov,masoliver1,masoliver2} for a more complete account of the CTRW formalism.

In the present work we will show two applications of CTRWs to the study of extreme problems in financial time series. We will take as underlying random process $X(t)$ the logarithmic price $X(t_n)=\ln(S(t_n))$, where $S(t)$ is the stock price at time $t$. We first consider the problem of obtaining the mean exit time of $X(t)$ out of a given interval $[a,b]$, of width $L$. We assume that at certain reference time $t_0$, right after an event, the price has a known value $X(t_0)=x_0$, $x_0 \in [a,b]$. Let us focus our attention on a particular realization of the process and suppose that at certain time $t_n>t_0$ the process first leaves the interval ---see Fig.~\ref{model}. We call the lapse $t_n-t_0$, the exit time out of the region $[a,b]$ and we will denote it by $t_{[a,b]}(x_0)$. This quantity is a random variable since it depends on the particular trajectory of $X(t)$ chosen and the MET is simply the average $T_{[a,b]}(x_0)=\text{E}[t_{[a,b]}(x_0)]$.

The standard approach to exit time problems is based on the knowledge of the survival probability ---see Sect.~\ref{Sect_SP}. In general, this is a quite involved path~\cite{weissrubin}. However, if we assume that $\tau_n$ and $\Delta X_n(\tau_n)$ are independent and identically distributed (i.i.d.) random variables, described by a joint pdf $\rho(x,\tau)$,
\begin{equation*}
\rho(x,\tau)dxd\tau=\text{Prob}\{x<\Delta X_n\leq x+dx;\tau<\tau_n\leq \tau+d\tau\}, 
\end{equation*}
it can be shown~\cite{montero1} that one can obtain the MET directly, without making use of the survival probability. In this framework the MET $T_{[a,b]}(x_0)$ obeys the following integral equation:
\begin{equation}
T_{[a,b]}(x_0)=\text{E}[\tau] +\int_a^bh(x-x_0)T_{[a,b]}(x)dx,
\label{met_iid}
\end{equation}
where $\text{E}[\tau]$ is the mean waiting time between jumps. It is worth noticing that Eq.~(\ref{met_iid}) is still valid even when $\tau_n$ and $\Delta X_n$ are cross-correlated. In fact, in the case of an i.i.d. process the MET only depends on $\rho(x,\tau)$ through its marginal pdfs $\psi(\tau)$ and $h(x)$.

We can illustrate the problem with a choice for $h(x)$, based on the small-scale properties of the system, which results in the observed~\cite{montero1,montero2} quadratic growth in the MET.
Let us introduce the following symmetrical two-state discrete model~\cite{montero2}:
\begin{equation}
h(x)=\frac{1}{2}\left[\delta(x-c)+\delta(x+c)\right].
\label{2D}                
\end{equation}
where $c$ is the basic jump size. This choice for $h(x)$ implies that the flat levels in every particular trajectory will be in a regular grid of size $c$ centred at the starting point $x_0$. It is worth noticing that this approach is also used in the context of option pricing, when the fair price of a derivative product is obtained by making use of the binomial trees methodology, where it is assumed that the stock price makes a jump up or down with some probability~\cite{CRR}. The solution of this problem, if we start from the middle of the interval, reads:
\begin{equation}
\frac{T_{[a,b]}(a+L/2)}{\text{E}[\tau]}=\left(1+\frac{L}{2 c}\right)^2. 
\label{met_iid_sol_exact_a}
\end{equation}
If we consider a symmetric exponential function for the jump distribution instead:   
\begin{equation}
h(x)=\frac{\gamma}{2}e^{-\gamma|x|},
\label{exponential}
\end{equation}
a very similar result is obtained \cite{montero1}:
\begin{equation}
\frac{T_{[a,b]}(a+L/2)}{\text{E}[\tau]}=\frac{1}{2}\left[1+\left(1+\frac{\gamma L}{2}\right)^2\right].
\label{exactmetsim}
\end{equation}

\section{Mean exit time for Markov-chain models}
\label{Sect_Markov}

In order to embrace also CTRWs with memory, we derived in~\cite{montero2} an integral equation for the MET 
when the jumps are Markovian. In particular, we focused on the case in which it is possible to neglect the  
influence of the past waiting time by
assuming that the magnitude of the previous change carries all the relevant information. 
The equation in this case is:
\begin{equation}
T_{[a,b]}(x_0|\Delta X_0)=\text{E}\left[\tau|\Delta X_0\right]
+ \int_{a}^{b} h(x-x_0|\Delta X_0)T_{[a,b]}(x|\Delta X) dx,
\label{MET_Markovian} 
\end{equation}
with $\Delta X=x-x_0$. Now the MET depends only on the marginal pdf of the return increments, $h(x|\Delta X_0)$, and on the conditional expectation of the waiting time, $\text{E}\left[\tau|\Delta X_0\right]$, which has to be evaluated through the marginal pdf, $\psi(\tau|\Delta X_0)$. In order to solve Eq.~(\ref{MET_Markovian}) and obtain explicit expressions for the MET we will use again a discrete two-state model:
\begin{equation*}
h(x|y) = \frac{c+ry}{2c} \delta(x-c)+\frac{c-ry}{2c} \delta(x+c),
\end{equation*}
where $r$ is the correlation between the magnitude of two consecutive jumps. The MET starting from the middle of the interval reads now:
\begin{equation*}
\frac{T_{[a,b]}(a+L/2)}{\text{E}[{\tau}]}=\frac{2r}{1+r}
\left(1+\frac{L}{2 c}\right)+\frac{1-r}{1+r}
\left(1+\frac{L}{2 c}\right)^2,
\end{equation*}
and, for large values of $L/c$, we recover the quadratic behaviour in the leading term:
\begin{equation*}
\frac{T_{[a,b]}(a+L/2)}{\text{E}[{\tau}]}\sim\frac{1-r}{1+r}\left(1+\frac{L}{2 c}\right)^2.
\end{equation*}

\section{The Survival Probability}\label{Sect_SP}

The survival probability is closely related to the MET as we will shortly show. It measures the likelihood that, up to time $t$, the process has been always in the interval $[a,b]$:
\begin{equation*}
 S_{[a,b]} (t - t_0 ;x_0 ) \equiv P\left\{ {a \le X(t) \le b,M(t) \le b,m(t) \ge a|X(t_0 ) = x_0 } \right\}, 
\end{equation*}
where we have defined the maximum and the minimum value of $X(t)$, $M(t)$ and $m(t)$, by:
\begin{equation*}
 M(t) = \mathop {{\rm{max}}}\limits_{t_0  \le t' \le t} {\rm{  }}X(t')\ {\rm{ ,  and   }}\ m(t) = \mathop {{\rm{min}}}\limits_{t_0  \le t' \le t} {\rm{ }}X(t'). 
\end{equation*}
The financial interest of SP is clear: it may be very useful in risk control. Note, for instance, the case $b\rightarrow \infty$. The SP measures, not only the probability that you do not loose more than $a$ at the end of your investment horizon, like VaR, but also in any previous instant. 

It is notorious that we can recover the MET from the Laplace Transform of the SP. If fact, as we have stated above, this is the standard technique used in the literature for obtaining METs. The link between both magnitudes becomes apparent if we express the MET in terms of $P\{ t_{[a,b]}  \le v|x_0 \}$, the cumulative distribution function (cdf) of the exit time:
\begin{eqnarray*}
 T_{[a,b]} (x_0) &=& \int_0^\infty  {vdP\{ t_{[a,b]}  \le v|x_0 \} }  =  \int_0^\infty  {\int_0^v {du{\rm{ }}dP} } \{ t_{[a,b]}  \le v|x_0 \}\\ 
  &=&  \int_0^\infty  {\int_v^\infty  {dP\{ t_{[a,b]}  \le v|x_0 \} du} }  =
 \int_0^\infty  {P\{ t_{[a,b]}  > u|x_0 \} du }.  
\end{eqnarray*}
Now, we must realize that the only way that $t_{[a,b]}$ can be bigger than any given value is that the process has been inside the interval up to that time: 
\begin{equation*}
P\left\{ {t_{[a,b]}  > t - t_0 |x_0 } \right\} =  \\ 
 P\left\{ {a \le X(t) \le b,M(t) \le b,m(t) \ge a|X(t_0 ) = x_0 } \right\},
\end{equation*}
and therefore,
\begin{equation*}
T_{[a,b]} (x_0 ) = \int_0^\infty  {S_{[a,b]} (u;x_0 )du}= \hat S_{[a,b]} (s = 0;x_0).
\end{equation*}

It is not surprising that the  
survival probability follows a renewal equation when also the mean exit time can be expressed in such a way ---see for instance Ref.~\cite{katja}. In the present case, where we consider that the process properties are depending, at most, on the size of last the jump, we can derive the following two-dimensional integral equation for the SP:
\begin{equation}
S_{[a,b]} (t - t_0 ;x_0 |\Delta X_0 ) = \Psi (t - t_0 |\Delta X_0 ) \\ 
  + \int_{t_0 }^t {dt'} \int_a^b {dx{\rm{  }}\rho (x - x_0 ,t' - t_0 |\Delta X_0 )S_{[a,b]} (t - t';x|\Delta X)}
\label{time_SP}
\end{equation}
where
\begin{equation*}
\Psi (t - t_0 |\Delta X_0 ) = \int_t^{ + \infty } {dt'} \int_{ - \infty }^{ + \infty } {dx{\rm{  }}\rho (x - x_0 ,t' - t_0 |\Delta X_0 )}
\end{equation*}
is the probability that the next sojourn will last more than $t-t_0$, given that the previous change was of size $\Delta X_0$. We can step down the dimension of the integral equation by considering the Laplace transform of Eq.~(\ref{time_SP}), 
\begin{equation*}
\hat S_{[a,b]} (s;x_0 |\Delta X_0 ) = \hat \Psi (s|\Delta X_0 ) + \int_a^b {dx{\rm{  }}\hat \rho (x - x_0 ,s|\Delta X_0 )\hat S_{[a,b]} (s;x|\Delta X)}.
\end{equation*}
Note that the problem is now much more complex, since it involves the joint pdf of jumps and sojourns, $\rho (x ,t|\Delta X_0 )$, not merely its marginal pdfs, $h(x|\Delta X_0)$ and $\psi(\tau|\Delta X_0)$. Even in the fully independent and case, the integral equation is hard to solve:
\begin{equation*}
\hat S_{[a,b]} (s;x_0) = \hat \Psi (s) +\hat \psi(s) \int_a^b {dx{\rm{  }} h (x - x_0 )\hat S_{[a,b]} (s;x)}.
\end{equation*}
The problem of the two-state discrete model without memory, Eq.~(\ref{2D}), is affordable but the complexity  of the solution casts few light into the general understanding of the issue. Therefore, we have left it for a forthcoming work, and we have focused our attention on the symmetric exponential case, Eq.~(\ref{exponential}), which gave similar results for the MET ---cf Eqs.~(\ref{met_iid_sol_exact_a}) and~(\ref{exactmetsim}). This model is very suitable for our purposes because reduces the problem from solving an integral equation to finding the solution of a second-order (ordinary) differential equation:
\begin{equation*}
\partial^2_{xx} \hat{S}_{[a,b]}(s;x)=\gamma^2(1-\hat{\psi}(s))\left[\hat{S}_{[a,b]}(s;x) -s^{-1}\right],
\end{equation*}
with the following boundary conditions:
\begin{equation*}
\partial_{x} \hat{S}_{[a,b]}(s;x=a)=\gamma\left[\hat{S}_{[a,b]}(s;a)-\hat{\Psi}(s)\right], \quad \partial_{x} \hat{S}_{[a,b]}(s;x=b)=-\gamma\left[\hat{S}_{[a,b]}(s;b)-\hat{\Psi}(s)\right]. 
\label{bc}
\end{equation*}
Even though, the final expression in the Laplace domain is so intricate:
\begin{equation*}
\hat{S}_{[a,b]}(s;x_0)=\frac{1}{s}\left[1-\hat{\psi}(s) \frac{\cosh \left\{\gamma \sqrt{1-\hat{\psi}(s)}\left(x_0-\frac{a+b}{2}\right)\right\}}{\sqrt{1-\hat{\psi}(s)}\sinh \left\{\gamma \sqrt{1-\hat{\psi}(s)}L/2\right\}+ \cosh\left\{ \gamma \sqrt{1-\hat{\psi}(s)}L/2\right\}}\right],
\end{equation*}
that, in general, it cannot be reverted to the time domain. The solution when the process begins at the center of the interval is somewhat simpler 
but still difficult to deal with:
\begin{equation*}
 \hat S_{[a,b]} (s;a + L/2) =  
 \frac{1}{s}\left[ {1 - \frac{{\hat \psi (s)}}{{\sqrt {1 - \hat \psi (s)} \sinh\left\{\gamma \sqrt {1 - \hat \psi (s)} L/2\right\} + \cosh\left\{ \gamma \sqrt {1 - \hat \psi (s)} L/2\right\}}}} \right].
\end{equation*}
The result when the interval width $L$ is infinite, but the process begins at finite distance of one of the boundaries, is even shorter:   
\begin{equation*}
\hat S_{( - \infty ,x]} (s;x_0 ) = \hat S_{[x,\infty )} (s;x_0 ) =  
\frac{1}{s}\left[ {1 - \frac{{\hat \psi (s)}}{{1 + \sqrt {1 - \hat \psi (s)} }}\exp\left\{ { - \gamma \sqrt {1 - \hat \psi (s)} |x - x_0 |} \right\}} \right],
\end{equation*}
and it can be directly compared with the same outcome for the Wiener process:
\begin{equation}
\hat S_{( - \infty ,x]} (s;x_0 ) = \hat S_{[x,\infty )} (s;x_0 ) = \frac{1}{s}\left[ {1 - \exp\left\{ { - {\frac{\sqrt{2s}}{\sigma}} |x - x_0 |} \right\}} \right], 
\label{SP_Inf}
\end{equation}
where the volatility $\sigma$ is the square root of the diffusion coefficient. The two formulas coincide for small values of the Laplace variable $s$, that is, for large timescales. The resemblance between both models when the interval width is bounded is not so evident, because in the Wiener case the SP can be only expressed in terms of an expansion series:
\begin{equation*}
\hat S_{[a,b]} (s;a + L/2) = 
\sum\limits_{k = 0}^\infty  {\frac{{8L^2 }}{{(2k + 1)\pi }}}  \cdot \frac{{( - 1)^k }}{{\sigma^2\pi ^2 (2k + 1)^2  + 2L^2 s}}.
\end{equation*}
In any case, it is easy to check that the long-term behaviour of the MET is similar: 
\begin{equation*}
T_{[a,b]} (a + L/2) = L^2 /4\sigma^2.
\end{equation*}

\section{Risk control} \label{Sect_Risk}
We will finally illustrate how SPs can be used in risk control. In order to clarify the concepts we will remove model-dependent inferences by using the outcome corresponding to the Wiener case. The Gaussian model is typically used for computing the ``Value of Risk" (VaR) level. VaR gives the worst return you can obtain at the end of a fixed time interval $t-t_0$, for a given confidence level $\alpha$. If we assume that the market volatility is $\sigma$, then
\begin{equation*}
\rm{VaR}=\sigma\sqrt{t-t_0} \NN^{-1}\left(1-\alpha\right),
\end{equation*}
where $\NN(\cdot)$ is the cdf for a Normal pdf. This measure of the risk exposure of an open position ignores the instantaneous risk aversion of the investor, since it neglects the fact that investors may not assume all the paths leading to the same final return. 

\begin{figure}[htbp]
\begin{center}
\includegraphics[width=0.90\textwidth,keepaspectratio=true]{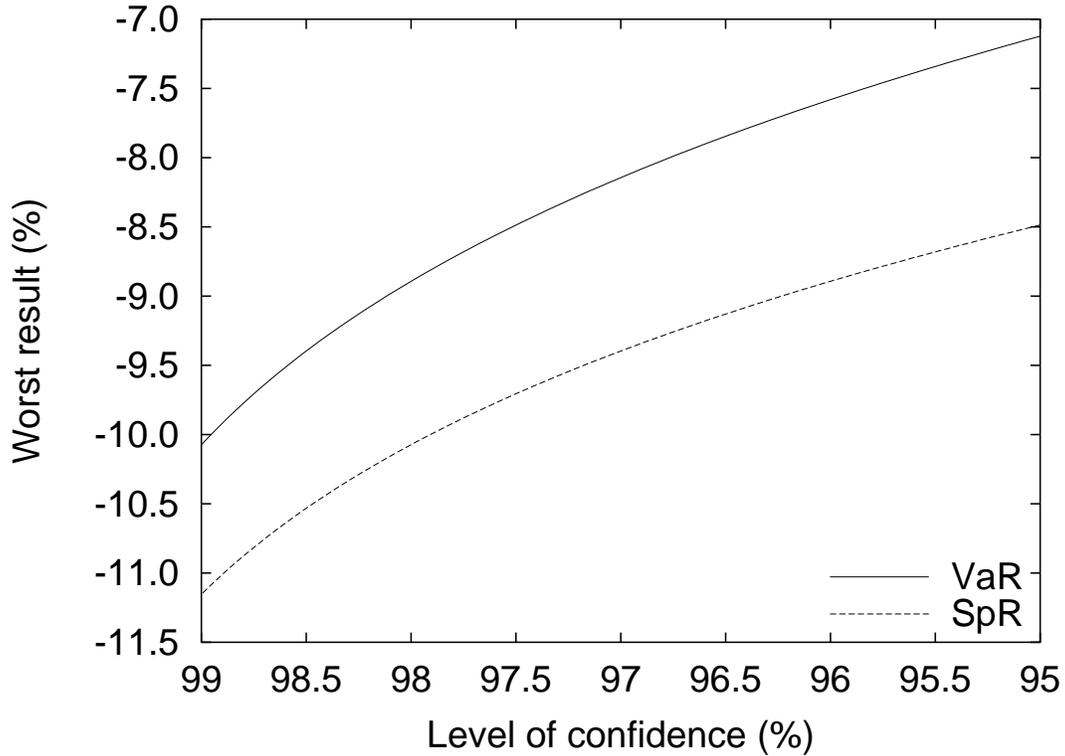} \caption{Risk values for a position lasting one month, if the process follows a Wiener process with a volatility of 15\%. We compare, for different confidence levels, the measure of the risk that both methods, VaR and SpR, yield. Clearly VaR underestimates the risk.} \label{VaR}
\end{center}
\end{figure}

This will not be the case if we use SP for quantifying the risk, since it will ensure, within the desired level of confidence, that the position is never below the risk measure, which we will call survival probability risk (SpR). The Laplace inverse transform on Eq.~(\ref{SP_Inf}) reads,
\begin{equation*}
S_{[x,\infty )} (t-t_0;x_0 ) = 1 - 2 \NN\left(  - {\frac{|x - x_0|}{\sigma\sqrt{t-t_0}}}  \right), 
\end{equation*}
and therefore
\begin{equation*}
\rm{SpR}=\sigma\sqrt{t-t_0} \NN^{-1}\left(\frac{1-\alpha}{2}\right).
\end{equation*}
In Fig.~\ref{VaR} we will found a comparative example with the two risk measures.

\section{Conclusions} \label{Sect_Conclusions}

We have argued for the convenience of the use of CTRWs in the modelling of stochastic processes in finance. CTRW is a well suited tool for representing market changes at very low scales, within the realm of high frequency data. We have shown that this formalism allows a thorough description of extreme events under a very general setting: we have obtained renewal integral equations for magnitudes related to these events when the return can be described by either an independent or a Makovian process. We have revisited the properties of the MET, a statistic that can inform about investment horizons. In previous works we found that it seems to scale in a similar way for different assets. We have addressed the topic of the SP in finance afterwards. SP has even more severe implications in risk management. SpR can improve the efficiency of more traditional methods, like VaR. We have introduced new theoretical results on this issue, and shown a practical example of its application.    

\acknowledgments The authors acknowledge support from MCyT under contract BFM2003-04574, and from MEC under contract FIS2006-05204.

\end{document}